\documentclass[preprint,aps,apl,superscriptaddress,amssymb,amsmath,floatfix,citeautoscript]{revtex4-1}
\usepackage[pdftex]{graphicx}
\usepackage{epstopdf}

\date{\today}

\begin{document}

\title{Crystalline Polymers with Exceptionally Low Thermal Conductivity Studied using Molecular Dynamics}

\author{Andrew B. Robbins}
\author{Austin J. Minnich}%
 \email{aminnich@caltech.edu}
\affiliation{%
 Division of Engineering and Applied Science\\
 California Institute of Technology, Pasadena, California 91125,USA
}%

\date{\today}

\begin{abstract}
	
	Semi-crystalline polymers have been shown to have greatly increased thermal conductivity compared to amorphous bulk polymers due to effective heat conduction along the covalent bonds of the backbone. However, the mechanisms governing the intrinsic thermal conductivity of polymers remain largely unexplored as thermal transport has been studied in relatively few polymers. Here, we use molecular dynamics simulations to study heat transport in polynorbornene, a polymer that can be synthesized in semi-crystalline form using solution processing. We find that even perfectly crystalline polynorbornene has an exceptionally low thermal conductivity near the amorphous limit due to extremely strong anharmonic scattering. Our calculations show that this scattering is sufficiently strong to prevent the formation of propagating phonons, with heat being instead carried by non-propagating, delocalized vibrational modes known as diffusons. Our results demonstrate a mechanism for achieving intrinsically low thermal conductivity even in crystalline polymers that may be useful for organic thermoelectrics.
	
\end{abstract}

\pacs{}
\maketitle 
\clearpage

Bulk polymers are generally considered heat insulators due to ineffective heat transport across the weak van der Waals bonds linking polymer chains. However, both computational\cite{zhang_polymer_2014,wang_thermal_2013,liu_length-dependent_2012,henry_high_2008,freeman_thermal_1987} and experimental\cite{wang_thermal_2013,shen_polyethylene_2010,choy_elastic_1999,mergenthaler_thermal_1992,piraux_thermal_1989,mugishima_phonon_1988} studies have demonstrated that some semi-crystalline polymers can have large thermal conductivity exceeding that of many metals. This enhancement occurs when the polymer chains are highly aligned, allowing heat to preferentially transport along the strong covalent backbone bonds. Thermally conductive polymers could find great use in a variety of heat dissipation applications including electronics packaging and LEDs.

Computational studies of heat transport in polymers have predicted the high thermal conductivity of crystalline polymers and also identified key molecular features that contribute to this high thermal conductivity. While most studies have focused on polyethylene (PE), comparisons of thermal conductivity among polymers have helped to identify factors of particular importance in setting a polymer's intrinsic upper limit to thermal conductivity. First, backbone bond strength\cite{zhang_polymer_2014} has been identified as heavily influencing group velocity, leading to higher thermal conductivity. Additionally, chain segment disorder\cite{luo_molecular_2011,zhang_morphology-influenced_2012,zhang_polymer_2014}, or the random rotations of segments in a chain, has been shown to lead to lower thermal conductivity. 

An important goal in exploiting the high intrinsic thermal conductivity of certain polymers is to fabricate semi-crystalline polymers at a large scale. A morphology potentially suited to this purpose is the polymer brush, or an array of polymer chains attached at one or both ends to a substrate\cite{losego_interfacial_2010}. A promising synthesis technique, known as surface-initiated ring-opening metathesis polymerization (SI-ROMP), is able to uniformly grow tethered polymer chains from a substrate in the desired aligned structure. Polynorbornene (PNb) and its derivatives are well studied in the ROMP synthesis technique and are thus of interest as thermal interface materials \cite{edmondson_polymer_2004}. However, the intrinsic thermal transport properties of PNb remain unknown.

In this Letter, we use molecular dynamics (MD) to study heat conduction in PNb. While PNb meets the standard criterion for high thermal conductivity, our simulations indicate that even perfectly crystalline PNb has an exceptionally low thermal conductivity nearly at the amorphous limit. We show that this low thermal conductivity arises from significant anharmonicity in PNb, causing high scattering rates that prevent the formation of phonons, resulting in heat being carried by non-propagating vibrations known as diffusons~\cite{allen_thermal_1993,allen_diffusons_1999}. Our work shows how intrinsically low thermal conductivity can be realized in fully crystalline polymers which may be of use for organic thermoelectrics.

\begin{figure*}[]
	\centering
	\includegraphics[width=6.5in]{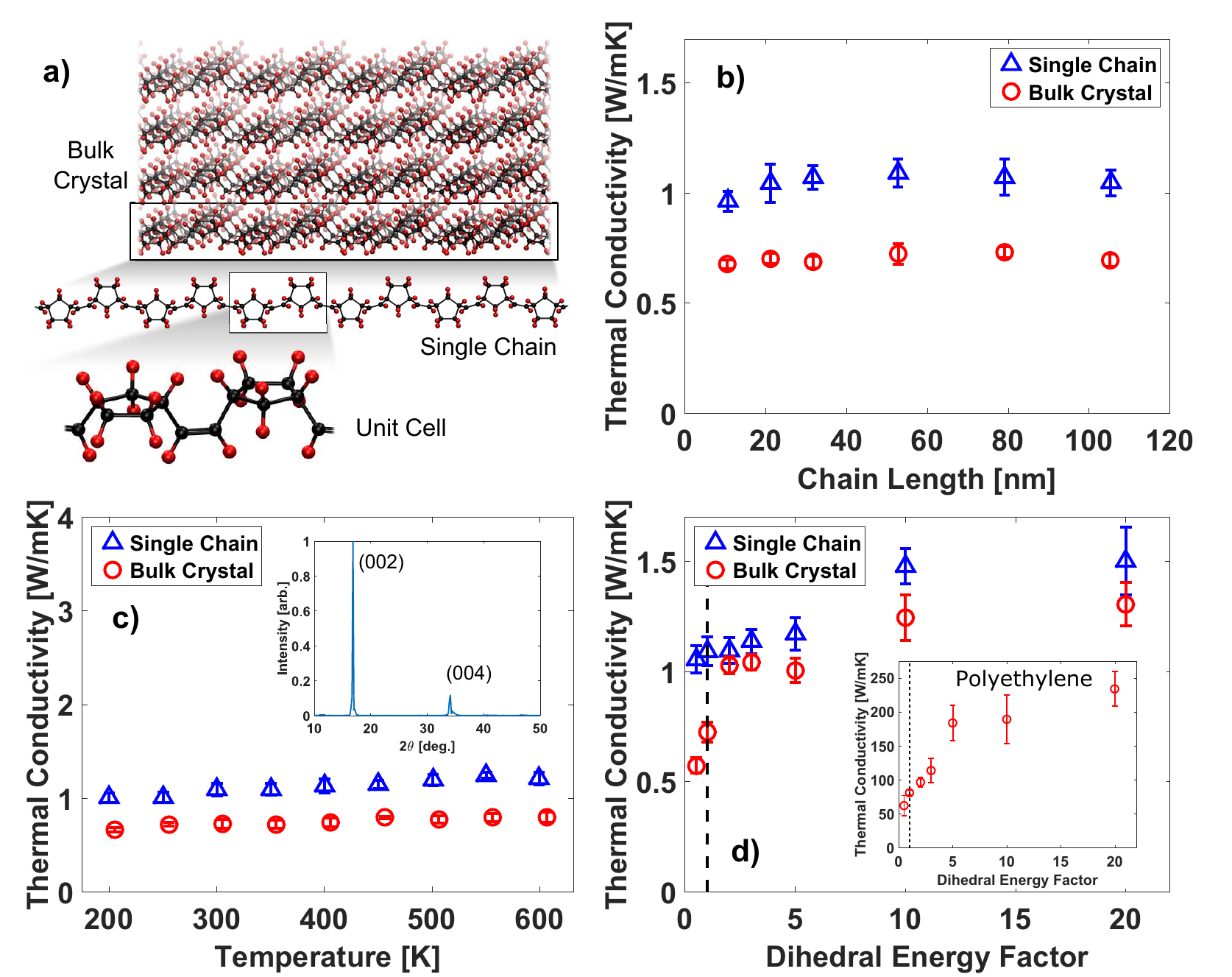}
	\caption{a) Illustration of a PNb crystal, a single isolated chain, and the unit cell, consisting of carbon (black) and hydrogen (red) atoms. b) Thermal conductivity as a function of chain length for both single chains (blue triangles) and bulk crystals (red circles), indicating thermal conductivity is independent of length. c) Thermal conductivity as a function of temperature for PNb single chains and crystals. Inset: Simulated diffraction pattern of PNb crystal at 300 K along the chain direction with peaks demonstrating crystallinity. The temperature dependence of the thermal conductivity is characteristic of amorphous materials despite the crystallinity of the polymer. d) Thermal conductivity as a function of a dihedral energy multiplication factor for both single chain and bulk crystal PNb. A larger factor represents a stiffer dihedral angle. The dotted line indicates a factor of 1, or no change to the potential. Inset: Identical calculation for a PE crystal. PNb demonstrates less dependence on dihedral energy compared to PE.}
	\label{fig:PNb_UC_TCvsTempandLenandDih}
\end{figure*}

We calculated the thermal conductivity of single chain and crystalline PNb using equilibrium MD simulations with the Large-scale Atomic/Molecular Massively Parallel Simulator (LAMMPS). The unit cell of PNb\cite{sakurai_crystal_1993} is shown in Fig.~\ref{fig:PNb_UC_TCvsTempandLenandDih}a and consists of 5-membered singly-bonded carbon rings each connected by a doubly-bonded carbon bridge. The simulation process begins by relaxing the PNb crystal in an NPT ensemble. The relaxed chain length obtained from the crystal is used for the isolated single chains of the same temperature. We use periodic boundary conditions in all directions. All simulations utilize the polymer consistent force field (PCFF).~\cite{sun_ab_1995}.

Once the structure is thermalized, we calculate the thermal conductivity using the Green-Kubo formalism,
\begin{equation}
	\kappa_z=\frac{V}{k_BT^2}\int_{0}^{\infty}\left\langle J_z(0)J_z(t)\right\rangle dt
\end{equation}
in an NVE ensemble. All simulations are run with 1 fs timesteps, and NVE ensembles are run for 3 ns to collect sufficient statistics for the thermal conductivity calculation. All thermal conductivity calculations in this work reflect the average of 5 or more identical simulations with different random initial conditions. Error bars reflect a single standard deviation of values across these simulations. 

The calculated thermal conductivity versus chain length at 300 K for both single chains and crystals is shown in Fig.~\ref{fig:PNb_UC_TCvsTempandLenandDih}b. We find the thermal conductivity of a simple bulk PNb crystal at 300 K along the chain direction is $0.72\pm .05$ W/mK and independent of length. All subsequent simulations use a chain length of 50 nm. For reference, the typical thermal conductivities of amorphous polymers are 0.1-0.4 W/mK\cite{mark_physical_2007}. The thermal conductivity perpendicular to the chain direction gives a thermal conductivity of $0.25\pm .03 $ W/mK, yielding an estimate of the lower limit to thermal conductivity mediated by weak inter-chain interactions. In contrast, polyethylene (PE) was found to have a thermal conductivity of $76.0\pm 7.2$ W/mK using the same procedure, in agreement with prior works.\cite{liu_length-dependent_2012} Thus, the thermal conductivity of crystalline PNb is very small and comparable to the amorphous value, in marked contrast to the large thermal conductivity of crystalline PE.

We also calculate the temperature dependent thermal conductivity of PNb, given in Fig.~\ref{fig:PNb_UC_TCvsTempandLenandDih}c. This result shows a weak, positive dependence of thermal conductivity on temperature, a feature characteristic of amorphous materials. This temperature dependence contrasts with the trend for crystals, exhibited by PE, where thermal conductivity decreases with temperature due to phonon-phonon scattering. Although the thermal conductivity of PNb has an amorphous trend, it remains fully crystalline. We verify the crystallinity of the structure by performing simulated diffraction on several snapshots of the crystal configuration at 300 K. As shown in the inset of Fig.~\ref{fig:PNb_UC_TCvsTempandLenandDih}c, clear peaks are observed, indicating that the structure remains crystalline even at room temperature.

To identify the origin of PNb's low thermal conductivity, we first consider the molecular features previously identified in the literature as influencing thermal conductivity. The first feature is, intuitively, strong backbone bonds. PNb has an entirely carbon backbone consisting of rings with double bonds between them. Compared to PE's purely single-bonded carbon backbone, PNb is composed of equal and stronger bonds, which would suggest a large group velocity and thus high thermal conductivity, contrary to our calculations.

The second criterion for high thermal conductivity is the absence of chain disorder, which breaks the translational symmetry of the crystal and causes scattering. PE, for example, has been shown to experience a dramatic drop in thermal conductivity at high temperatures where the repeating units begin to chaotically rotate with respect to one another.\cite{zhang_morphology-influenced_2012} The onset of such disorder is connected with the rotational stiffness of the chain, which arises directly from the magnitude of the dihedral angle energy terms in the interatomic potential.~\cite{sun_ab_1995} Stiffer dihedral terms prevent rotation around a given bond in the molecule and thus inhibit segmental rotation. 

To quantitatively relate the effect of this stiffness with heat transport, we artificially alter the dihedral terms in the potential and observe its effect on thermal conductivity. Figure~\ref{fig:PNb_UC_TCvsTempandLenandDih}d shows this result for both an isolated PNb chain and a PNb crystal. The inset shows the results of identical calculations done on PE crystals. For PNb, both single chain and bulk crystal show a weak dependence on stiffness. In contrast, PE shows a large increase in thermal conductivity as the dihedral energy is increased. These observations suggest that chain disorder does not play a significant role in PNb, and certainly cannot explain its exceptionally low thermal conductivity. This conclusion is further supported by previous studies of amorphous PNb that find it has highly restricted rotational movement along the chain.~\cite{haselwander_polynorbornene:_1996}

\begin{figure*}[]
	\centering
	\includegraphics[width=6.5in]{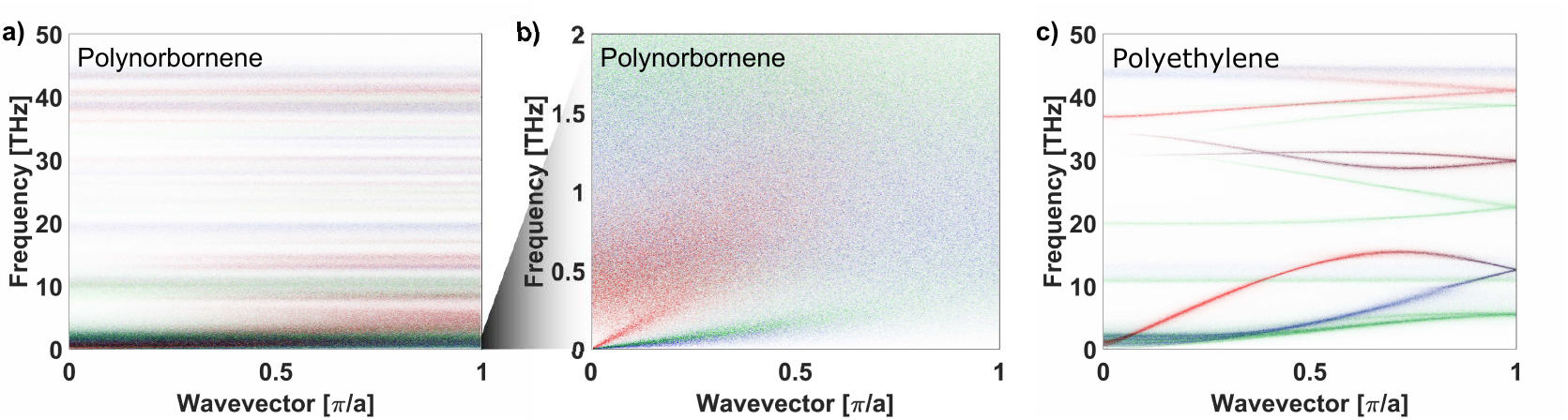}
	\caption{a) Phonon dispersion for crystalline PNb at 300 K calculated from MD, showing contributions from motion in the x(blue), y(green), and z(red) directions with the chains aligned in the z direction. b) Zoomed in phonon dispersion from (a). Longitudinal acoustic (red) and transverse acoustic (blue/green) modes are clearly visible at low frequencies. c) Phonon dispersion for crystalline polyethylene at 300 K computed identically as in (a). To avoid the dispersion being washed out by high intensities from low frequencies, intensities are normalized by a certain percentile of brightness. (a) and (b) are normalized to the 96\% and (c) to the 99\%. The log of the intensities are used in (a) and (c) to better show the whole dispersion.}
	\label{fig:PNbandPE_disp}
\end{figure*}

To identify the origin of PNb's low thermal conductivity, we next find the phonon dispersion for PNb by calculating the spectral energy density as a function of frequency, $\omega$, and wavevector, $k_z$,~\cite{thomas_predicting_2010,zhang_polymer_2014,feng_anharmonicity_2015}
\begin{equation}
	\Phi_\alpha(\omega,k_z)=C\sum_{a=1}^{N_a}\left|\int\sum_{n=1}^{N_z}v_\alpha(z_{n},t)e^{i(k_zz_{n}-\omega t)}dt\right|^2
\end{equation}
where $C$ is a constant, $\alpha=\{x,y,z\}$ selects the velocity component along the specified axis, $n$ indexes the unit cell up to $N_z$ unit cells in the chain, and $a$ indexes the backbone atoms within a unit cell up to $N_a$ backbone atoms. Figure~\ref{fig:PNbandPE_disp}a shows the intensity plots of $\Phi_x(\omega,k_z)$, $\Phi_y(\omega,k_z)$, and $\Phi_z(\omega,k_z)$, equivalent to the phonon dispersion, calculated for a PNb crystal at 300 K shown in blue, green, and red, respectively. Polymer chains are aligned in the z direction. The most striking feature of the dispersion is the absence of well-defined modes, with a blurred, diffuse background instead of crisp phonon bands. Figure~\ref{fig:PNbandPE_disp}b shows a zoomed in view of PNb's dispersion, enabling clear longitudinal acoustic (LA) modes in red and transverse acoustic modes (TA) in blue and green to be observed below approximately 0.25 THz. 

In contrast, the dispersion for crystalline PE at 300 K calculated in Fig.~\ref{fig:PNbandPE_disp}c shows crisp phonon bands throughout the frequency range,  in agreement with the literature \cite{braden_inelastic_1999,tomkinson_inelastic_2002,barrera_vibrational_2006}. The transition from clearly discernible modes at low frequencies to poorly-defined modes at high frequencies seen in Fig.~\ref{fig:PNbandPE_disp}b for PNb has been observed before in other quasi-1D structures~\cite{chen_twisting_2015}, though not in polymers, and is known as the Ioffe-Regel crossover.~\cite{ioffe_non-crystalline_1960} The crossover occurs when phonon mean free paths (MFPs), $\ell$, are comparable to the phonon wavelength, $\lambda$. If the thermal vibration scatters after traveling only the order of a wavelength, the vibration cannot be described as a propagating wave and hence defining a wavevector is not possible.

To confirm whether scattering prevents the formation of phonons, we calculated the phonon lifetimes and MFPs from the phonon dispersion using frequency-domain normal mode analysis by fitting the linewidth of a chosen polarization in the dispersion to a Lorentzian.~\cite{thomas_predicting_2010,feng_anharmonicity_2015} For a particular polarization, $j$, and wavevector, $k_z$, the Lorentzian fitting yields the full width at half max $2\Gamma_{k_z,j}$ which is related to the mode lifetime by $\tau^{-1}=2\Gamma_{k_z,j}$.

The full dispersion, $\Phi(\omega,k_z)$, contains contributions from all polarizations, but because the LA modes correspond solely to atomic motion in the z direction at low frequencies, as evidenced by the isolated red modes in Fig.~\ref{fig:PNbandPE_disp}b, $\Phi_{LA}(\omega,k_z)=\Phi_z(\omega,k_z)$ for frequencies less than 0.25 THz. All subsequent fittings represent results for the LA polarization only.

\begin{figure*}[]
	\centering
	\includegraphics[width=6.5in]{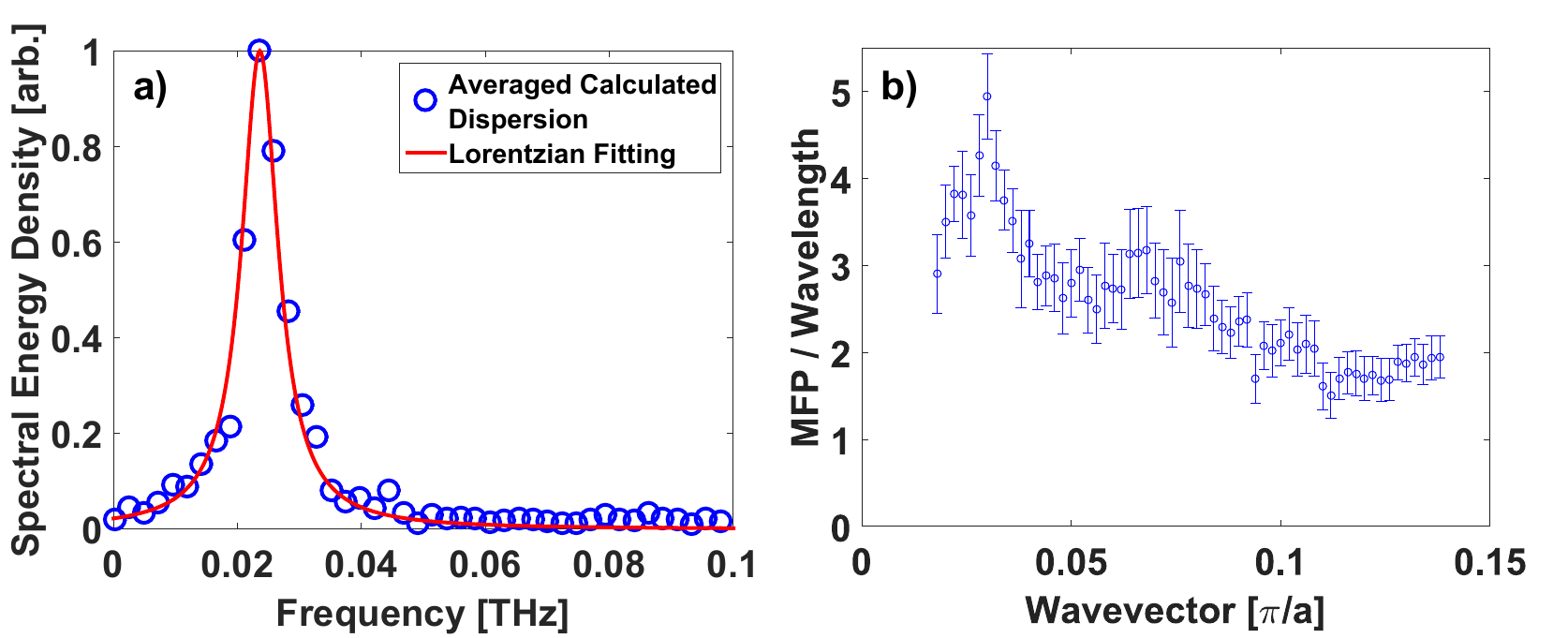}
	\caption{a) Spectral energy density, $\Phi_{LA}(\omega,k_z)$, for the longitudinal acoustic polarization shown in Fig.~\ref{fig:PNbandPE_disp}b at $k_z=0.02\left(\frac{\pi}{a}\right)$ as a function of frequency. The data points represent an average across nearby wavevectors and the red line represents the Lorentzian fitting that yields the lifetimes shown in (b). b) Mean free paths, $\ell$, normalized by wavelength, $\lambda$, for the longitudinal acoustic modes of crystalline polynorbornene at 300 K as a function of wavevector. Error bars represent the 95\% confidence interval according to the fitting. The Ioffe-Regel crossover occurs when $\ell \sim \lambda$.}
	\label{fig:lifetimes}
\end{figure*}

The Lorentzian fitting was repeated at all wavevectors spanned by the LA polarization. The peaks were then fit to obtain a smooth dispersion curve, $\omega(k_z)$, with an average group velocity $v_g = 3050$ m/s. To achieve a better signal to noise ratio, the values of $\Phi_{LA}(\omega,k_z)$ at closely adjacent wavevectors were averaged after matching their peaks. The data centered at $k_z=0.02\left(\frac{\pi}{a}\right)$ and its fitting are shown in Fig.~\ref{fig:lifetimes}a. 

The resulting lifetimes give the MFPs, $l=v_g\tau$ and are plotted, normalized by wavelength, $\lambda$, in Fig.~\ref{fig:lifetimes}b as a function of wavevector. The figure clearly shows that the condition for Ioffe-Regel crossover, $l\sim\lambda$, becomes satisfied as wavevector is increased. At higher values of $k_z$ than are shown in the figure, fitting the linewidth of the dispersion becomes impossible due to the absence of a well-defined peak. Correspondingly, looking at the full dispersion in Fig.~\ref{fig:PNbandPE_disp}b, it is clear that beyond a frequency of approximately 0.25 THz, the discernible LA modes no longer exist and become ill-defined.

These calculations demonstrate that the exceptionally low thermal conductivity in crystalline PNb is the direct result of the absence of propagating vibrations to carry heat caused by strong scattering. Instead, heat is carried by non-propagating, delocalized thermal vibrations known as diffusons~\cite{allen_thermal_1993,allen_diffusons_1999}. This explanation is consistent with the absence of well-defined modes in the dispersion as well as the characteristic amorphous temperature dependence observed in the thermal conductivity of PNb despite its crystalline structure.

Finally, we provide additional evidence for the strong anharmonicity present in PNb by calculating the Gruneisen parameter $\gamma$, defined as
\begin{equation}
	\gamma = -\frac{\partial\ln\omega}{\partial\ln V}
	\label{eq:gruneisen}
\end{equation}
where $\omega$ is the phonon mode frequency and $V$ is the system volume. While $\gamma$ varies for different phonon modes and depends on temperature, we can estimate it by calculating the phonon dispersion for a range of temperatures and observing the change in volume and phonon frequencies. We restrict this calculation to the same LA branch used to calculate the lifetimes and for frequencies below the Ioffe-Regel crossover, where the modes are defined. Three dispersions were calculated at 12 temperatures between 260 K and 400 K. The PNb crystals were relaxed at each temperature to obtain the system volume and dispersions were fit to obtain $\omega(k_z,T)$ over the wavevector range of the LA mode. At each wavevector $k_z$, data for $\ln[\omega(k_z,T)]$ vs $\ln [V]$ were linearly fit to obtain an effective $\gamma$. The value obtained for PNb after averaging over these modes is -2.8. Identical calculations for PE over its LA mode yield $\gamma$ = 0.044, significantly smaller in magnitude compared to PNb. Due to the large anisotropy of the polymer crystal, it may also be useful to consider $L_z^3$ as the relevant effective volume rather than $V$ in Eq.~\ref{eq:gruneisen}. The average $\gamma$ values for PNb and PE are -3.9 and -0.32, respectively. The difference in values by more than an order of magnitude remains. Also note $\gamma$ becomes negative for PE due to the negative coefficient of linear thermal expansion in the chain direction. The negative value for $\gamma$ for PNb is likely due to the complicated atomic structure and the complex molecular force field, which strongly differ from typical crystals where bond stretching is more commonly associated with phonon softening. These results strongly suggest that strong anharmonicity is present in PNb and is responsible for the large intrinsic scattering rates.

In summary, we have used MD simulations to show that crystalline PNb has an exceptionally low thermal conductivity near the amorphous limit due to strong anharmonic scattering that prevents the formation of phonons, and thus possesses intrinsically low thermal conductivity. While PNb is not suitable for thermal management applications, the mechanism for low thermal conductivity identified here could be of interest for organic thermoelectric applications as perfectly crystalline polymers may have good charge carrier mobility but retain low thermal conductivity. This work highlights the wide range of thermal transport properties that can be realized in polymers, which may prove useful in employing polymers in thermal applications.

\section*{Acknowledgments}

This work was supported by an ONR Young Investigator Award under Grant Number N00014-15-1-2688 and by startup funds from Caltech. The authors thank Professor Bill Goddard and members of his group for assistance with the LAMMPS software.

\clearpage
\bibliographystyle{is-unsrt}
\bibliography{MyLibrary}

\end{document}